\documentclass[prd,notitlepage,twocolumn,nofootinbib,
               superscriptaddress,preprintnumbers]{revtex4-1}
\pdfoutput=1
\usepackage[english]{babel}
\usepackage{amsmath,amssymb,amsfonts, bm,bbm,slashed, subdepth}
\usepackage{graphicx}
\usepackage[sort&compress]{natbib}
\usepackage{xcolor}
\usepackage[normalem]{ulem}
\usepackage{hyperref}
\usepackage{cleveref}
\definecolor{red}{rgb}{1.0, 0, 0}
\usepackage{hyperref}
\usepackage{enumerate}
\usepackage{epsfig, subfigure}
\usepackage{setspace}
\usepackage{booktabs, tabularx}
\usepackage{units}
\usepackage{hyperref}

\newcommand{\be}{\begin{equation}}
\newcommand{\ee}{\end{equation}}
\newcommand{\ba}{\begin{array}}
\newcommand{\ea}{\end{array}}
\newcommand{\bea}{\begin{eqnarray}}
\newcommand{\eea}{\end{eqnarray}}
\newcommand{\balg}{\begin{align}}
\newcommand{\ealg}{\end{align}}
\newcommand{\bit}{\begin{itemize}}
\newcommand{\eit}{\end{itemize}}
\newcommand{\trm}[1]{\textrm{#1}}

\newcommand{\Mpc}{\trm{\Mpc}}
\newcommand{\yr}{\trm{\yr}}
\newcommand{\eV}{\trm{\eV}}

\newcommand{\ev}[1]{\ensuremath{\langle #1 %
                     \rangle}} 

\begin{document}

\title{Sterile Neutrinos with Secret Interactions --- Lasting Friendship with Cosmology}

\author{Xiaoyong Chu}
\email{xchu@ictp.it}
\affiliation{International Center for Theoretical Physics,
             Strada Costiera 11, 34014 Trieste, Italy.}

\author{Basudeb Dasgupta}
\email{bdasgupta@theory.tifr.res.in}
\affiliation{Tata Institute of Fundamental Research,
             Homi Bhabha Road, Mumbai, 400005, India.}

\author{Joachim Kopp}
\email{jkopp@uni-mainz.de}
\affiliation{PRISMA Cluster of Excellence and Mainz Institute for Theoretical Physics,
             Johannes Gutenberg University, 55099 Mainz, Germany.}

\preprint{MITP/15-033, TIFR/TH/15-14}

\begin{abstract}
Sterile neutrinos with mass $\simeq 1$~eV and order 10\% mixing with active
neutrinos have been proposed as a solution to anomalies in neutrino oscillation
data, but are tightly constrained by cosmological limits. It was recently shown
that these constraints are avoided if sterile neutrinos couple to a new MeV-scale
gauge boson $A'$. However, even this scenario is restricted by structure
formation constraints when
$A'$-mediated collisional processes lead to efficient active-to-sterile
neutrino conversion after neutrinos have decoupled. In view of this, we
reevaluate in this paper the viability of sterile neutrinos with such ``secret''
interactions.  We carefully dissect their evolution in the early Universe,
including the various production channels and the expected modifications to
large scale structure formation.  We argue that there are two regions in
parameter space --- one at very small $A'$ coupling, one at relatively large $A'$
coupling --- where all constraints from big bang nucleosynthesis (BBN), cosmic
microwave background (CMB), and large scale structure
(LSS) data are satisfied. Interestingly, the large $A'$ coupling region is precisely the 
region that was previously shown to have
potentially important consequences for the small scale structure of dark matter
halos if the $A'$ boson couples also to the dark matter in the Universe.
\end{abstract}

\maketitle

\section{Introduction}
\label{sec:intro}

The possible existence of extra, ``sterile'', neutrino species with masses at
the eV scale and $\mathcal{O}(10\%)$ mixing with the Standard Model (SM)
neutrinos is one of the most debated topics in neutrino physics today.  Several
anomalies in neutrino oscillation experiments~\cite{Aguilar:2001ty,
AguilarArevalo:2012va, Mueller:2011nm, Mention:2011rk, Hayes:2013wra,
Acero:2007su} seem to point towards the existence of such particles, but null
results from other experiments that did not observe a signal cast doubt on this
hypothesis~\cite{Kopp:2011qd, Kopp:2013vaa, Conrad:2012qt, Kristiansen:2013mza,
Giunti:2013aea}.  A multi-faceted experimental program is under way to clarify
the issue and either detect, or conclusively rule out, eV-scale sterile
neutrinos with large mixing angle.

If the SM is indeed augmented with one or several such sterile neutrinos, but
nothing else, some of the tightest constraints come from cosmological
observations.  In particular, measurements of the effective number of
relativistic particle species in the primordial plasma,
$N_\text{eff}$~\cite{Steigman:2012ve,Planck:2015xua}, disfavor the existence of
an abundance of light or massless particles beyond the SM neutrinos and the
photon in the early Universe.  If sterile neutrinos are at the eV scale or above,
they are also constrained by the distribution of large scale structure
(LSS) in the Universe~\cite{Hamann:2011ge} 
which would be washed out due to
efficient energy transport over large distances by free-streaming neutrinos.

Cosmology, however, only constrains particle species that are abundantly
produced in the early Universe.  In two recent letters~\cite{Hannestad:2013ana,
Dasgupta:2013zpn}, it was demonstrated that the production of sterile neutrinos
can be suppressed until relatively late times if they are charged under a new
interaction. This idea has elicited interest in detailed model building and has
interesting phenomenological consequences\,\cite{Bringmann:2013vra, Ko:2014nha,
Ng:2014pca, Kopp:2014fha, Saviano:2014esa, Mirizzi:2014ama, Cherry:2014xra,
Kouvaris:2014uoa,Tang:2014yla}. However, Mirizzi \emph{et al.}~\cite{Mirizzi:2014ama} 
have recently pointed out that collisions mediated by the new interaction can
result in significant late time production of sterile neutrinos and lead to
tensions with CMB and LSS data on structure formation.
There are, however, several important caveats to this statement. In particular,
the bounds from Ref.~\cite{Mirizzi:2014ama} can be evaded if the sterile neutrinos
either never recouple with active neutrinos or remain collisional until
matter-radiation equality. (We communicated on these caveats with the authors of
Ref.~\cite{Mirizzi:2014ama}, who were aware of the first possibility
but did not mention it as they found it less interesting in the 
context of previous literature. They mention the second possibility in the
final version of their paper.)

Our aim in the present paper is to understand in detail the role and impact of
sterile neutrino collisions, and reevaluate if secretly interacting sterile
neutrinos remain cosmologically viable. We begin in Sec.~\ref{sec:setup} by
reviewing the main features of self-interacting sterile neutrino scenarios.
Then, in Sec.~\ref{sec:Neff}, we calculate the additional contribution to
$N_{\rm eff}$ at the BBN and CMB epochs.
In Sec.~\ref{sec:LSS} we consider the impact on the
large scale structure in the Universe, focusing in particular
on the sensitivity
of Sloan Digital Sky Survey (SDSS) and Lyman-$\alpha$ data. We find that there
are two regions of
parameter space where sterile neutrinos with secret interactions are only
weakly constrained. In Sec.~\ref{sec:conclusions}, we discuss our conclusions
and summarize the results.

\section{Secret Interactions and Sterile Neutrino Production }
\label{sec:setup}

We assume that the Standard Model is augmented by a sterile neutrino $\nu_s$ with mass
$m_s$\footnote{Since $\nu_s$ is not a mass eigenstate, $m_s$ actually means the
mass of the fourth, mostly sterile, mass eigenstate.} and with order 10\% mixing
with the SM neutrinos.  We moreover assume the existence of a new secret
$U(1)_{s}$ gauge interaction, mediated by a vector
boson $A'$ of mass $M$ at the MeV scale and coupling to sterile neutrinos through an
interaction of the form
\begin{align}
  {\cal L}_\text{int} = e_s \bar{\nu}_s \gamma^\mu P_L \nu_s A'_\mu \,.
\end{align}
Here, $e_s$ is th $U(1)_s$ coupling constant and $P_L = (1 - \gamma^5)/2$ is
the left-handed chirality projection operator.
We define the secret fine structure constant as $\alpha_s \equiv e_s^2/(4\pi)$.

This new interaction generates a large temperature-dependent
potential~\cite{Dasgupta:2013zpn}
\begin{align}
  V_\text{eff} \simeq
  \left\lbrace 
    \begin{array}{lcl}
      -\dfrac{7 \pi^2 e_s^2 E T_s^4}{45 M^4}  &\quad& \text{for $T_s \ll M$} \\[3ex]
      +\dfrac{e_s^2 T_s^2}{8 E}               &\quad& \text{for $T_s \gg M$}
    \end{array}
  \right.\,
  \label{eq:Veff}
\end{align}
for sterile neutrinos of energy $E$ and sterile sector temperature $T_s$.
This potential leads to an in-medium mixing angle $\theta_m$ between active neutrinos $\nu_a$
and sterile neutrinos $\nu_s$, given by
\begin{align}
  \sin^2 2\theta_m
    = \frac{\sin^2 2\theta_0}
           {\big(\cos 2\theta_0 + \tfrac{2 E}{\Delta m^2} V_\text{eff} \big)^2
           + \sin^2 2\theta_0} \,.
  \label{eq:s22thm}
\end{align}
In the following, we will use a vacuum mixing angle $\theta_0 \simeq 0.1$
and an active--sterile mass squared difference $\Delta m^2 \simeq 1\,\text{eV}^2$.
As shown in \cite{Hannestad:2013ana, Dasgupta:2013zpn},
the secret interactions can suppress $\theta_m$, and
thus active to sterile neutrino oscillations, until after neutrino decoupling
as long as $|V_{\rm eff}|\gg |\Delta m^2/(2E)|$.

The new interaction also leads to collisions of sterile neutrinos. The
collision rate for $\nu_s\nu_s\leftrightarrow \nu_s\nu_s$ scattering is given
by
\begin{align}
\Gamma_\text{coll}= n_{\nu_s} \sigma\sim 
  \left\lbrace 
    \begin{array}{lcl}
      n_{\nu_s} e_s^4\frac{E^2}{M^4}
                           &\quad& \text{for $T_s \ll M$} \\[1.5ex]
      n_{\nu_s} e_s^4\frac{1}{E^2}
                           &\quad& \text{for $T_s \gg M$}
    \end{array}
  \right. \,,
  \label{eq:Gamma-coll}
\end{align}
where $n_{\nu_s}$ is the sterile neutrino density. The sterile neutrino
production rate $\Gamma_s$ and the final density depend on this collision
rate. Two qualitatively different scenarios must be distinguished:

\emph{Collisionless production:} If the collision rate $\Gamma_\text{coll}$ is
smaller than the Hubble rate $H$ at all times, the active and sterile neutrinos
can be taken to be oscillating without
scattering~\cite{Dodelson:1993je}.\footnote{We ignore the SM matter potential
and scattering experienced by active neutrinos because we will be interested in
the regime where the secret interaction dominates over the SM interaction.}
If $\Delta m^2/(2 T_{\nu_a}) \gg H$, $\nu_s$ are then produced only
through oscillations, so that the final
sterile neutrino number density is $n_{\nu_s}\simeq\tfrac{1}{2} \sin^2
2\theta_m\,n_{\nu_a}$, where
$n_{\nu_a}=3\zeta(3)/(4\pi^2)g_{\nu_a}T_{\nu_a}^3$ is the density of one of
active neutrino flavors and $T_{\nu_a}$ is the active neutrino temperatur.
The final population of sterile neutrinos thus
remains small, at most ${\cal O}(10^{-2})$ of the active neutrino
density, because of the small mixing angle.

\emph{Collisional production:} If $\Gamma_\text{coll}$ exceeds the Hubble rate
$H$, then sterile neutrinos cannot be treated as
non-collisional~\cite{Stodolsky:1986dx}. In each collision, the sterile
component of a $\nu_a$--$\nu_s$ superposition changes its momentum, separates from
the $\nu_a$ component, and continues to evolve independently. Subsequently, the
active component again generates a sterile component, which again gets
scattered.  This process continues for all neutrinos until eventually the phase
space distributions of $\nu_a$ and $\nu_s$ have become identical. Thus, the
fraction of $\nu_a$ converted to sterile neutrinos is not limited by the mixing
angle, and all neutrino flavors end up with equal number densities.

The $\nu_a\rightarrow \nu_s$ production rate in this case is $\Gamma_s \simeq
\frac{1}{2}\sin^2 2\theta_m \cdot \Gamma_\text{coll}$~\cite{Jacques:2013xr}, where
we can interpret the first factor as the average probability that an initially active
neutrino is in its sterile state at the time of collision.  The second factor
gives the scattering rate that keeps it in the sterile state. We note that the
production rate $\Gamma_s$ is proportional to $n_{\nu_s}$ and thus rapidly
approaches its final value, 
\begin{equation}
  \Gamma_s \simeq {1\over2}\sin^2 2\theta_m \times
    \frac{3}{4} n^\text{SM}_{\nu_a} \cdot
    \left\lbrace 
      \begin{array}{lcl}
        e_s^4\frac{E^2}{M^4}
                             &\quad& \text{for $T_s \ll M$} \\[1.5ex]
        e_s^4\frac{1}{E^2}
                             &\quad& \text{for $T_s \gg M$}
      \end{array}
    \right. \,.
  \label{eq:Gamma-s}
\end{equation}
Note that, when $\Gamma_\text{coll}$ is much larger than the oscillation
frequency, using the average oscillation probability $\frac{1}{2}\sin^2
2\theta_m$ is inappropriate, and in fact the production rate $\Gamma_s$ goes to
zero in this case.  Such a situation is, however, not realized for the
parameter values explored in this work.

In the following, we will look at both collisionless and collisional production 
of sterile neutrinos in more detail,\footnote{There is also the possibility that
Mikheyev-Smirnov-Wolfenstein (MSW) type resonant effects, e.g., because of the
sign-flip of the secret potential $V_\text{eff}$ around $T_s \simeq M$, modify the $\nu_s$
production probability. In this work we treat all MSW transitions to be
completely non-adiabatic and thus ignore them.  A careful momentum-dependent
treatment, which we defer to future work, is needed to accurately describe resonant
conversion.} with a special focus on the latter where more sterile 
neutrinos may be produced.

\section{Constraints on $N_\text{eff}$}
\label{sec:Neff}

Cosmology is sensitive to the presence of relativistic degrees of freedom
through their contribution to the overall energy density.  At early times the
sterile sector was presumably in equilibrium with the SM plasma, so that
$\nu_s$ and $A'$ were thermally populated.  We assume that the sterile sector
decouples from the SM sector well above the QCD scale and that oscillations
remain suppressed until active neutrinos also decouple. Since the temperature
$T_\gamma$ of the SM sector drops more slowly than the sterile sector temperature
$T_s$ when extra entropy is produced during the QCD phase transition,
$T_s$ at BBN is significantly smaller than $T_\gamma$.

It is useful to track the ratio
\begin{align}
  \xi \equiv \frac{T_s}{T_{\nu}^\text{SM}}
\end{align}
of the sterile sector temperature $T_s$ and the the temperature
$T_\nu^\text{SM}$ of a standard neutrino.  Before $e^+e^-$ annihilation,
$T_{\nu}^\text{SM}
= T_\gamma$, while afterwards $T_{\nu}^\text{SM} = (4/11)^{1/3} T_\gamma$.
Assuming comoving entropy is conserved, the ratio $\xi$ at BBN is
\begin{align}
  \xi_\text{BBN}
    &= \left\lbrace 
         \begin{array}{ll}
           \big( \frac{10.75}{106.7} \big)^\frac{1}{3}
             \big( \frac{2 \cdot 7/8 + 3}{2 \cdot 7/8} \big)^\frac{1}{3}
              &\quad\text{for $M \gg 0.5$~MeV} \\[1ex]
           \big( \frac{10.75}{106.7} \big)^\frac{1}{3}
              &\quad\text{for $M \ll 0.5$~MeV}
         \end{array} \right. \nonumber\\[1ex]
    &= \left\lbrace
         \begin{array}{lcl}
           0.649  &\quad\text{for $M \gg 0.5$~MeV} &\quad\text{(case A)} \\[1ex]
           0.465  &\quad\text{for $M \ll 0.5$~MeV} &\quad\text{(case B)}
         \end{array} \right.\,.
  \label{eq:xi-BBN}
\end{align}
Here, the factor $(10.75/106.7)^{1/3}$ gives the ratio of the sterile sector
temperature to the active sector temperature before $A'$ decay, assuming
that the two sectors have decoupled above the electroweak scale. It is based
on counting the SM degrees of freedom that freeze out between the electroweak and
BBN epochs. 

$A'$ is present in the Universe at the BBN epoch if 
$M \ll 3T_s|_{\rm BBN}\simeq0.5$~MeV, and has decayed away if heavier. The factor $(2 \cdot 7/8 + 3) / (2 \cdot 7/8)$ in the
first row of Eq.~\eqref{eq:xi-BBN} corresponds to the ratio of sterile sector
degrees of freedom\footnote{Note that in complete
models, for instance in scenarios including a dark Higgs sector to break the $U(1)_s$
symmetry, more degrees of freedom may need to be taken into account in the above
equations.} before and after the decay of $A'$ at 
$T_s\simeq M/3$.

The extra radiation in the Universe is parameterized as
$N_\text{eff} \equiv (\rho_{\nu_a} + \rho_{\nu_s,A'}) / \rho_\nu^\text{SM}$,
i.e., the energy density of all non-photon relativistic species, measured in units of the
energy density of a SM neutrino species.  The primordial population of $\nu_s$
and $A'$ leads to $N_{\rm eff}$ marginally larger than 3. 
For $M \gg 0.5\ \text{MeV}$,
\begin{subequations}
\renewcommand{\theequation}{\theparentequation\Alph{equation}}
\begin{align}
  N_\text{eff,BBN{\tiny (A)}}
    &= N_{\nu_a} + \xi_\text{BBN{\tiny (A)}}^{4}
    \simeq 3.22 \,,
  \label{eq:N-BBN1}
\end{align}
at the BBN epoch.
The first term, $N_{\nu_a}$, on the right hand side accounts for the active 
neutrinos and is equal to 3.045.
The second term includes the relativistic sterile sector particles, i.e., only
$\nu_s$ if $M \gg 0.5\ \text{MeV}$.
If the $A'$ bosons are lighter, i.e., $M \ll 0.5\ \text{MeV}$, they
are present during BBN and contribute $g_{A'} = 3$ degrees
of freedom in the sterile sector, in addition to the $g_{\nu_s} =
2 \times 7/8$ degrees of freedom of a sterile neutrino. Using the fact that
also each active neutrino species has $g_{\nu_a} = 2 \times 7/8$ degrees of
freedom, we find
\begin{align}
  N_\text{eff,BBN{\tiny (B)}}
    &= N_{\nu_a} + \frac{g_{\nu_s} + g_{A'}}{g_{\nu_a}} \xi_\text{BBN{\tiny (B)}}^{4}
     \simeq 3.17 \,,
    \label{eq:N-BBN}
\end{align}
\end{subequations}

\begin{figure}
  \centering
  \includegraphics[width=0.95\columnwidth]{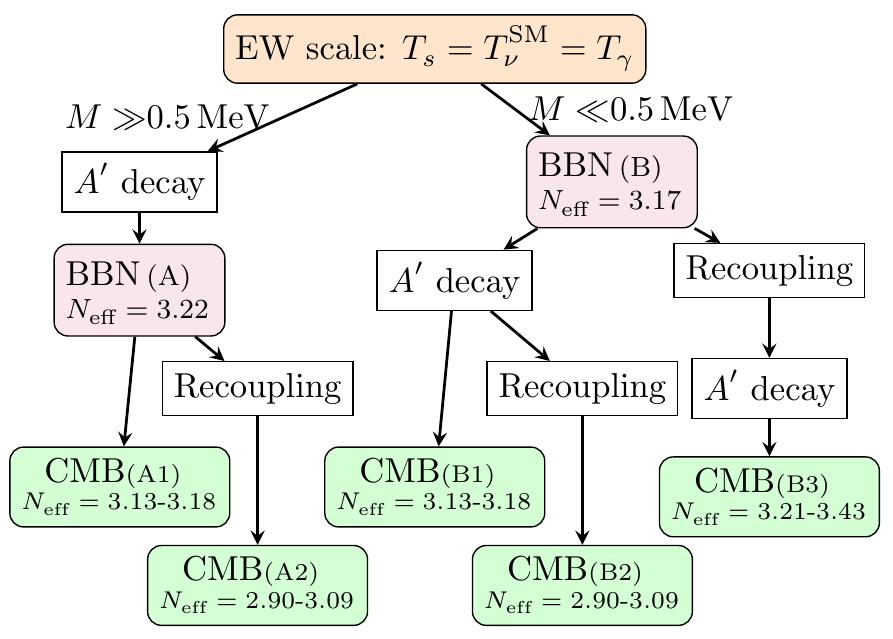}
  \caption{Possible cosmological histories of the active neutrinos $\nu_a$, the
    sterile neutrinos $\nu_s$, and the sterile sector gauge bosons $A'$ below the
    electroweak (EW) scale. Various possibilities, labeled as A1, A2 and
    B1, B2, B3, are determined by the values of the $A'$ mass $M$ and the $U(1)_s$
    fine structure constant $\alpha_s$ and lead to testable predictions for
    $N_\text{eff}$ at both the BBN and CMB epochs. See text for details.}
  \label{fig:chart}
\end{figure}

In Fig.~\ref{fig:chart}, these two
cases are summarized  as BBN\,{(A)} and BBN\,{(B)},
respectively. In either case, $N_\text{eff,BBN{\tiny (A/B)}}$ remains
consistent with the current BBN bound on extra radiation,
$\Delta N_\text{eff,BBN} = 0.66^{+0.47}_{-0.45}$ (68\%\,C.L.)~\cite{Steigman:2012ve}.

After BBN, the next important event is a possible secret \emph{recoupling} of
$\nu_a$ and $\nu_s$. If the sterile neutrino production rate $\Gamma_s > H$, a 
new hotter population of $\nu_s$ can be collisionally produced from $\nu_a$, 
and they achieve kinetic equilibrium with the primordially produced colder 
population of $\nu_s$. Also, the $A'$ can decay and heat up the sterile neutrinos. 
Both processes change the number and energy density of neutrinos, and 
$N_{\rm eff}$ at CMB depends on the order in which they occur.

\begin{figure}
  \centering
  \includegraphics[width=0.9\columnwidth]{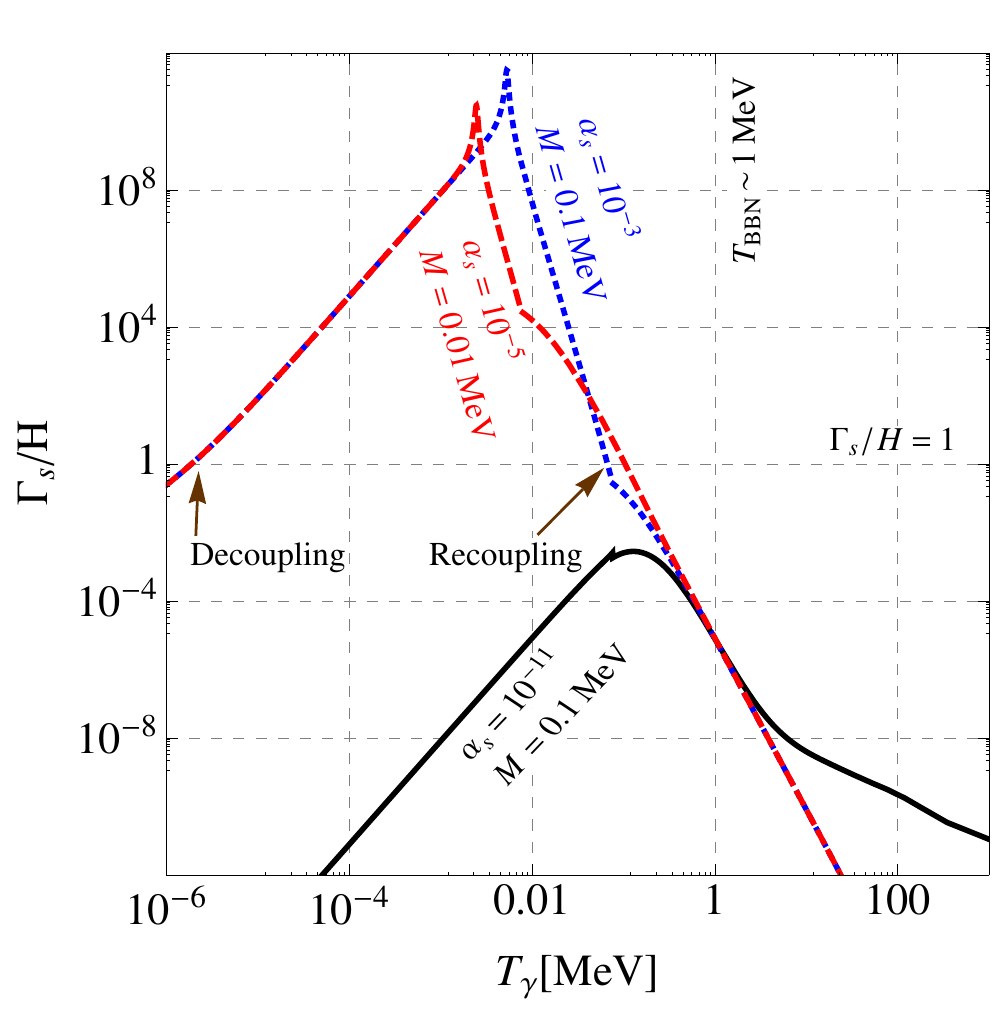}
  \caption{Evolution of the collisional $\nu_a\rightarrow \nu_s$ production
    rate $\Gamma_s$, normalized
    by the Hubble rate $H$, versus the photon temperature $T_\gamma$, for
    different representative choices of the secret gauge boson
    mass $M$ and the
    secret fine structure constant $\alpha_s$. When $\Gamma_s/H > 1$,
    collisional production of $\nu_s$ from the thermal bath of $\nu_a$ is
    effective.  The solid black curve shows a case where this never happens.
    The shoulder around $T_\gamma \sim M$ is where the $A'$ decay away. The dotted blue
    and dashed red curves correspond to recoupling after and before $A'$ decay,
    respectively.}
  \label{fig:thermalization-rate}
\end{figure}

In Fig.~\ref{fig:thermalization-rate}, we show the
collisional $\nu_a\rightarrow \nu_s$ production rate $\Gamma_s$, normalized to the Hubble expansion rate $H$, as
a function of the photon temperature $T_\gamma$:
$\Gamma_s/H$ is suppressed at high temperatures (say, above GeV), where $\sin^2
2\theta_m$ is small due to the large $V_\text{eff}$. Since $\Gamma_s \propto
T_\gamma^{-3}$ in this regime (see Eqs.~\eqref{eq:Veff}, \eqref{eq:s22thm} and
\eqref{eq:Gamma-s}) and $H \propto T_\gamma^2$, $\Gamma_s/H$ increases with
$T_\gamma^{-5}$ as the temperature decreases. We define the
{recoupling} temperature $T_\text{re}$ as the temperature where $\Gamma_s
/H > 1$ for the first time since the primordial decoupling of the active and
sterile sectors above the QCD phase transition. When $T_s \sim M$, the energy
and temperature dependence of $\Gamma_s$ changes (see Eq.~\eqref{eq:Gamma-s}),
and when also $V_\text{eff}$ drops below $\Delta m^2 / (2 T_s)$ at
$T_s < M$, $\Gamma_s$ begins to drop again. The asymptotic behavior is
$\Gamma_s / H \propto T_\gamma^3$ at $T_s \ll M$ and $\theta_m \simeq
\theta_0$. There are then three possible sequences of events:
\begin{enumerate}
  \item \emph{No recoupling}: For a sufficiently small interaction strength
    $\alpha_s$, the scattering rate $\Gamma_s$ always stays below the
    Hubble rate and there is no recoupling (solid black curve in
    Fig.~\ref{fig:thermalization-rate}).
\end{enumerate}
If the interaction is stronger, a recoupling of $\nu_a$ and $\nu_s$ can happen either 
\emph{after} or \emph{before} $A'$ decay:
\begin{enumerate}
\setcounter{enumi}{1}
  \item \emph{Recoupling after $A'$ decay}: If $M > \text{few} \times
    10^{-2}$~MeV, the recoupling happens after $A'$ have decayed (dotted blue
    curve in Fig.~\ref{fig:thermalization-rate}).

  \item \emph{Recoupling before $A'$ decay}: If $M < \text{few} \times
    10^{-2}$~MeV, the recoupling happens before $A'$ have decayed (dashed red
    curve in Fig.~\ref{fig:thermalization-rate}).
\end{enumerate} 
In the second and third cases, there is also a secret \emph{decoupling} when
$\Gamma_s/H$ again drops below one.  If $e_s^2/M^2 \le {\cal O}(10\ \text{MeV}^{-2})$,
this decoupling happens while $\nu_s$ are still relativistic, 
i.e.\ $T_s \gtrsim m_s/3$.\footnote{In this paper, we will always assume this to
be the case since we will find that the parameter region with $e_s^2/M^2 \ge {\cal
O}(10\ \text{MeV}^{-2})$ is already disfavored by the requirement that active
neutrinos should free stream sufficiently early~\cite{Cyr-Racine:2013jua} (see
Secs.~\ref{sec:LSS} and \ref{sec:conclusions}). If $\nu_s$ and $\nu_a$
are still coupled when the $\nu_s$ become non-relativistic, the mostly sterile
mass eigenstate $\nu_4$ will undergo
a non-relativistic freeze-out and partly annihilate to pairs of
mostly active neutrinos.
Similarly, there is the possibility that the $A'$ decay after the decoupling, 
but this does not happen for the range of parameters we will discuss here.}
 
In the following, we discuss the three aforementioned cases in detail.

\subsubsection{No Recoupling}

In the \emph{no recoupling} cases, labeled as A1 and B1 in Fig.~\ref{fig:chart},
the cosmological evolution after BBN is very straightforward. Vacuum oscillations
convert a small fraction $\simeq \frac{1}{2} \sin^2 2\theta_0 \simeq 0.01$ of
active neutrinos into sterile neutrinos (and vice versa), but 
this has negligible impact on the cosmological observables. Therefore, the
temperature ratio $\xi$
at CMB can be derived from the separate conservation of entropy in the active neutrino
sector and in the sterile neutrino sector. It is independent of when the $A'$
decay (provided that it happens before the CMB epoch and approximately in chemical equilibrium.)
That is, $\xi_\text{CMB\tiny{A/B}} \simeq \xi_\text{BBN\tiny{(A)}} = 0.649$. 

$N_\text{eff,CMB}$ can be estimated in analogy to Eq.~\eqref{eq:N-BBN1}.  For the
assumed sterile neutrino mass $\simeq 1$~eV, the $\nu_s$
contribution to the
relativistic energy density has to be weighted by an extra factor because they
are already semi-relativistic at the CMB epoch, where the photon temperature is
$T_\gamma \simeq 0.30$~eV and the kinetic temperature of the sterile sector is
$T_s = \xi_\text{CMB}\cdot (4/11)^{1/3} T_\gamma \simeq
0.14$~eV. As in \cite{Mirizzi:2014ama},
we assume that the extra weight factor is characterized by the pressure
$P$. (See Appendix~\ref{sec:Tkin} for the definition and calculations of the kinetic
temperature and the pressure $P$ used here.) We thus obtain
\begin{align}
  N_\text{eff,CMB}
    &= N_{\nu_a} + \frac{P_{m_s=\text{1\,eV}}}{P_{m_s=0}}
                        \Bigg|_\text{CMB} \cdot  \xi_\text{CMB}^{4}
    \simeq 3.13 \,.
  \label{eq:N-CMB1}
\end{align}
It is worth noting that the CMB temperature spectrum does not exactly measure
the value of $N_\text{eff,CMB}$. Instead, the observed spectrum depends on the
evolution of the energy density in relativistic degrees of freedom between the
epoch of matter-radiation equality ($T_{\gamma,\text{eq}} \sim 0.7$\,eV) and
recombination ($T_{\gamma,\text{CMB}} \simeq 0.30$~eV)~\cite{Planck:2015xua}.
Therefore, the value of $N_\text{eff,CMB}$ measured from the CMB
temperature power spectrum lies between the values of $N_\text{eff}$ at
$T_{\gamma,\text{CMB}}$ and $T_{\gamma,\text{eq}}$. The latter value, which we
will denote by $N_\text{eff,eq}$, is given by
\begin{align}
  N_\text{eff,eq}
  &= N_{\nu_a} + \frac{P_{m_s=\text{1\,eV}}}
                             {P_{m_s=0}}
                        \Bigg|_\text{eq} \cdot  \xi_\text{eq}^{4} \simeq 3.18 \,.
\end{align}
Both of these values agree with the bound from the 2015 Planck data release,
$N_\text{eff} = 3.15 \pm 0.23$\ (68\% C.L.)~\cite{Planck:2015xua}.

\subsubsection{Recoupling after $A'$ decay}

The cases of \emph{recoupling after $A'$ decay} are labeled as A2 and B2 in
Fig.~\ref{fig:chart}. In both cases, entropy conservation in the sterile sector
before recoupling leads to a temperature ratio just after $A'$ decay of
$\xi_M \simeq \xi_\text{BBN\tiny{(A)}} = 0.649$, which in turn implies 
\begin{align}
  N_{\text{eff},M} = 3.045 + \xi^4_M = 3.22 \,.
\end{align}
Here, we have assumed that during $A'$ decay chemical equilibrium holds in
the sterile sector.

After recoupling, efficient neutrino oscillations and collisions
lead to equilibration of the
number densities and energy densities of all active and sterile neutrino
species. Nevertheless, since number-changing interactions are strongly
suppressed at $T_s \ll M$, this recoupling
cannot change the total (active + sterile) neutrino
number density and energy density beyond what is necessitated by cosmological
expansion. The kinetic temperature $T_{\nu,\text{re}}$ shared by all neutrinos
after recoupling is then given by
\begin{align}
  T_{\nu,\text{re}} \simeq
    \frac{3.045 \cdot (T^\text{SM}_{\nu,\text{re}})^4 + 1 \cdot T_{s,\text{re},0}^4}
         {3.045 \cdot (T^\text{SM}_{\nu,\text{re}})^3 + 1 \cdot T_{s,\text{re},0}^3}
  \simeq 0.97 \, T^\text{SM}_\nu \,,
\end{align}
where $T^\text{SM}_\nu$ is the active neutrino temperature just prior
to recoupling (which is at its SM value) and $T_{s,\text{re},0}$ denotes the
sterile sector temperature just prior to recoupling.

Eventually, the mostly sterile eV-scale mass eigenstate decouples from
the light mass eigenstates and becomes semi-relativistic at the CMB time.
Its kinetic temperature at this epoch is $T_{s,\text{CMB}} \simeq 0.13$~eV.
The effective number of relativistic species at the CMB epoch
is given by
\begin{align}
  N_\text{eff,CMB} &=
    N_\text{eff,$M$} \bigg( \frac{3}{4}
      + \frac{1}{4} \frac{P_{m_s=\text{1\,eV}}}{P_{m_s=0}}
      \bigg|_\text{CMB}
    \bigg) \notag \\
  &\simeq 2.51 + 0.39 \simeq 2.90 \,.
\end{align}
Note that this is smaller than the SM
value 3.045. This happens because part of the energy of active
neutrinos has been transferred to the mostly sterile mass eigenstate
$\nu_4$, whose kinetic energy gets redshifted away more efficiently after it
becomes non-relativistic.
Ref.~\cite{Mirizzi:2014ama} also found $N_\text{eff} < 3$ for this
scenario.  Similarly, we obtain for the time of matter-radiation equality:
\begin{align}
  N_\text{eff,eq} = 3.09 \,.
\end{align}
Both values are in agreement with the Planck bound~\cite{Planck:2015xua}.

\subsubsection{Recoupling before $A'$ decay}

The last possibility, labeled as case B3 in Fig.~\ref{fig:chart}, is that
recoupling happens before $A'$ decay.  In this case, all neutrinos, together
with $A'$, reach a common chemical equilibrium, which lasts until most of the
$A'$ particles have decayed. During the formation of chemical equilibrium, the
total energy is conserved while entropy increases. Energy conservation
allows us to calculate the temperature $T_{\nu,\text{re}}$ of the active + sterile
neutrino sector immediately after recoupling:
\begin{multline}
 (3 g_{\nu_a} + g_{\nu_s} + g_{A'}) \, T_{\nu,\text{re}}^4 \\
    = \big[ 3 g_{\nu_a} + (g_{\nu_s} + g_{A'}) \,
      \xi_\text{BBN{\tiny (B)}}^4 \big] \, (T^{SM}_{\nu,\text{re}})^4 \,,
\end{multline}
where $T^{SM}_{\nu,\text{re}}$ is again the active neutrino temperature
just prior to recoupling.
Plugging in numbers for the effective numbers of degrees of freedom
$g_{\nu_a}$, $g_{\nu_s}$, $g_{A'}$ and using $\xi_\text{BBN{\tiny (B)}}
= 0.465$, we obtain
\begin{align}
  T_{\nu,\text{re}} = 0.861 \, T^{SM}_{\nu,\text{re}} \,.
\end{align}
Later, the $A'$ decay and the thermal bath of neutrinos gets reheated 
by a factor $[(3 g_{\nu_a} + g_{\nu_s} + g_{A'}) / (3 g_{\nu_a} + g_{\nu_s})]^{1/3}
\simeq 1.125$.  The effective number of relativistic species after $A'$ decay is
then
\begin{align}
  N_{\text{eff},M} \simeq
    \frac{3 g_{\nu_a} + g_{\nu_s}}{g_{\nu_a}} \,
    (1.125 \cdot 0.861)^4 \simeq 3.568.
\end{align}
The next steps are the decoupling of sterile neutrinos and active neutrinos,
and then the freeze-out of sterile neutrino self-interactions.  Since the
number densities and energy densities of the different species do not change
during these decouplings, the total effective number of relativistic degrees of freedom
at the CMB epoch is given by
\begin{align}
  N_\text{eff,CMB} &=
    N_\text{eff,$M$} \bigg( \frac{3}{4}
      + \frac{1}{4} \frac{P_{m_s=\text{1\,eV}}}{P_{m_s=0}}
      \bigg|_\text{CMB}
    \bigg)
  \simeq 3.21 \,,
\end{align}
where again the pressure characterizes the contribution of the semi-relativistic
$\nu_4$ to the radiation density in the Universe.  Similarly, at
matter-radiation equality we have
\begin{align}
  N_\text{eff,eq} & \simeq 3.43 \,.
\end{align}
This number is still within the $2\sigma$ error of the Planck bound~\cite{Planck:2015xua}.

\section{Structure Formation}
\label{sec:LSS}

Besides the constraints on extra radiation measured by $N_\text{eff}$, CMB data
also prefers that most of the active (massless) neutrinos start to free-stream
before redshift $z \sim 10^5$~\cite{Archidiacono:2013dua, Cyr-Racine:2013jua}.
On the other hand, matter power spectrum observations forbid these
free-streaming degrees of freedom from carrying so much energy as to suppress
small scale structures~\cite{Lesgourgues:2012uu}. Therefore measurements of the
matter power spectrum put the most stringent upper bound on the mass of all
fully\ thermalized neutrino species: $\sum m_\nu \lesssim 0.2$--$0.7$~eV (95\%
C.L.)~\cite{Planck:2015xua}.  This concern~\cite{Mirizzi:2014ama} excludes a
large proportion of the parameter region for self-interacting sterile neutrinos
considered in \cite{Dasgupta:2013zpn}.  However, like the constraint on
$N_\text{eff}$ discussed in Sec.~\ref{sec:Neff}, it is avoided if $\Gamma_s$
never exceeds $H$ after the epoch when $V_\text{eff}$ drops below the
oscillation frequency (cases A1 and B1 in Fig.~\ref{fig:chart}).

Interestingly, structure formation constraints are \emph{also} significantly
relaxed when the $U(1)_s$ gauge coupling $e_s$ is large and/or the gauge
boson mass $M$ is small. In this case, sterile neutrinos, although produced
abundantly through collisional production (see Sec.~\ref{sec:setup}), cannot
free-stream until late times, long after matter-radiation equality. Thus, their
influence on structure formation is significantly reduced. We will now discuss
this observation in more detail.

After the active and sterile neutrinos have equilibrated through $A'$-mediated
collisions, they should be treated as an incoherent mixture of the four mass
eigenstates $\nu_i$ ($i = 1\dots4$). The reason is that for $m_4 \sim 1$~eV,
their oscillation time scales are much smaller than both the Hubble time and
the time interval between scatterings.
For simplicity, assume that only the mostly sterile mass
eigenstate $\nu_4$ is massive with mass $m_4 \simeq 1$~eV, and that it only
mixes appreciably with one of the mostly active mass eigenstates, say $\nu_1$:
\begin{align}
  \nu_s \simeq \sin\theta_0 \,\nu_1 + \cos \theta_0 \,\nu_4 \,.
\end{align}
We take the vacuum mixing angle to be $\theta_0 \simeq 0.1$ and we take into account
that matter effects are negligible at temperatures
relevant for structure formation (after matter--radiation equality).
Since it is the flavor eigenstate
$\nu_s$ that is charged under $U(1)_s$, the mass eigenstates $\nu_1$ and $\nu_4$
interact with relative rates $\sin^2\theta_0$ and $\cos^2\theta_0$, respectively,
while $\nu_2$ and $\nu_3$ essentially free-stream.

To study the influence of the secret interaction on structure formation, we
estimate the mean comoving distance $\lambda_s$ that each $\nu_4$ can
travel in the early Universe.  Since neutrinos can transport energy efficiently
over scales smaller than $\lambda_s$, the matter power spectrum will be
suppressed on these scales.  As long as neutrinos are collisional, they do not
free stream, but diffuse over scales of order~\cite{Kolb:1988aj}
\begin{align}
  (\lambda_s^\text{coll})^2 \simeq \int_0^{t_s^\text{dec}} \! dt \,
    \frac{\ev{v_s}^2}{a^2(t)} \, \frac{1}{n_s\,\ev{\sigma v}_s} \,,
  \label{eq:lambda-Silk}
\end{align}
where $a(t)$ is the scale factor of the Universe, $t_s^\text{dec}$ is the time
at which sterile neutrino self-interactions decouple,
\begin{align}
  \ev{\sigma v}_s \sim \ev{v_s}
    \frac{e_s^4 \cos^2\theta_0}{(M^2 + T_s^2)^2} \, (T_s + m_s)^2
  \label{eq:sigma-v}
\end{align}
is the thermally averaged interaction cross section of the mostly sterile mass
eigenstate $\nu_4$, estimated here by na\"ive dimensional analysis, and
$n_s$ is the number density of sterile neutrinos.
For simplicity, we take the kinetic temperature $T_s$ of the sterile sector
equal to the active neutrino temperature in this section, i.e.\
$T_s = T_\nu^\text{SM} = (4/11)^{1/3} T_\gamma \propto a^{-1}(t)$, as long as
$\nu_4$ are relativistic.  If sterile neutrinos become non-relativistic ($T_s <
m_s$) while they are still strongly self-coupled, the kinetic temperature
of the sterile sector scales as $T_s \propto a^{-2}(t)$ until $T_s$ drops below
$T_{s,\text{dec}}$.  After that, the sterile neutrino momenta are simply
redshifted proportional to $a^{-1}(t)$.
This implies in particular that, at $T_s \gg m_s$,
we have $n_s \simeq T_s^3$, while after $\nu_4$ become non-relativistic,
but are still strongly coupled, this changes to $n_s \propto T_s^{3/2}$.
The computation of the average velocity
$\ev{v_s}$ of $\nu_4$ entering eq.~\eqref{eq:sigma-v} is discussed in
Appendix~\ref{sec:Tkin}.

The decoupling temperature $T_{s,\text{dec}}$ and the corresponding time
$t_s^\text{dec}$ are defined by the condition that
the sterile neutrino interaction rate is just equal to the Hubble rate:
\begin{align}
  n_s \, \ev{\sigma v}_s \big|_{t=t^\text{dec}}
    = H(t^\text{dec}) \,.
  \label{eq:fs-cond}
\end{align}
After $t^\text{dec}$, sterile neutrinos start to free stream.  The total
comoving distance that a $\nu_4$ travels between the time $t^\text{dec}$ and
the present epoch $t^0$ is~\cite{Kolb:1988aj}
\begin{align}
  \lambda^\text{fs}_s
    = \int_{t^\text{dec}}^{t_0} \! dt \, \frac{\ev{v_s(t)}}{a(t)} \,.
  \label{eq:lambda}
\end{align}
The overall damping scale is then given by
\begin{align}
  \lambda_s^2  =  (\lambda_s^\text{coll})^2  +  (\lambda^\text{fs}_s)^2
\end{align}
At scales larger than $\lambda_s$, structure formation is unaffected by
the existence of sterile neutrinos, while at smaller scales, structures are
washed out.

\begin{figure*}[!t]
  \centering
  \begin{tabular}{cc}
    \includegraphics[width=0.95\columnwidth]{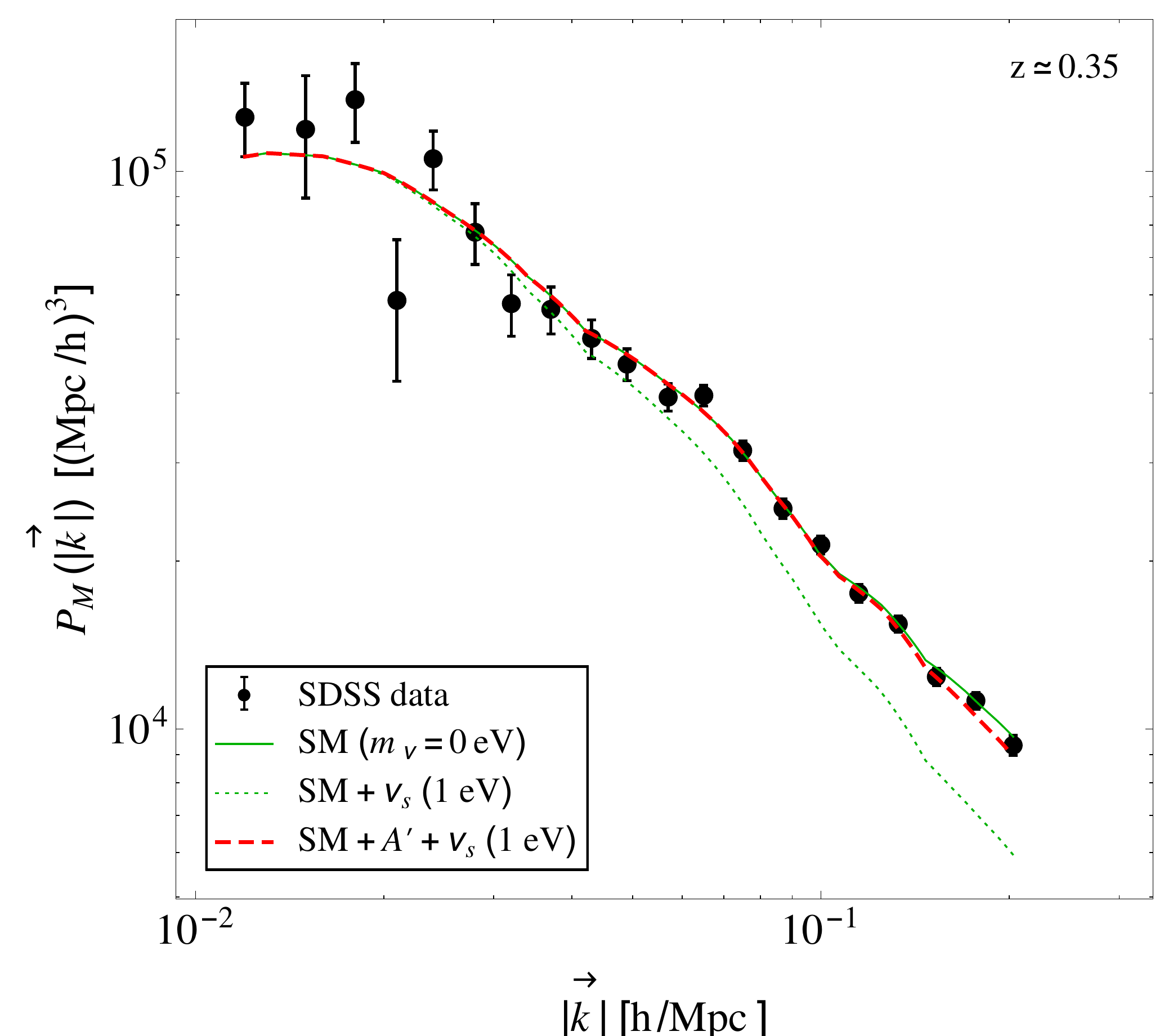} &
    \includegraphics[width=0.95\columnwidth]{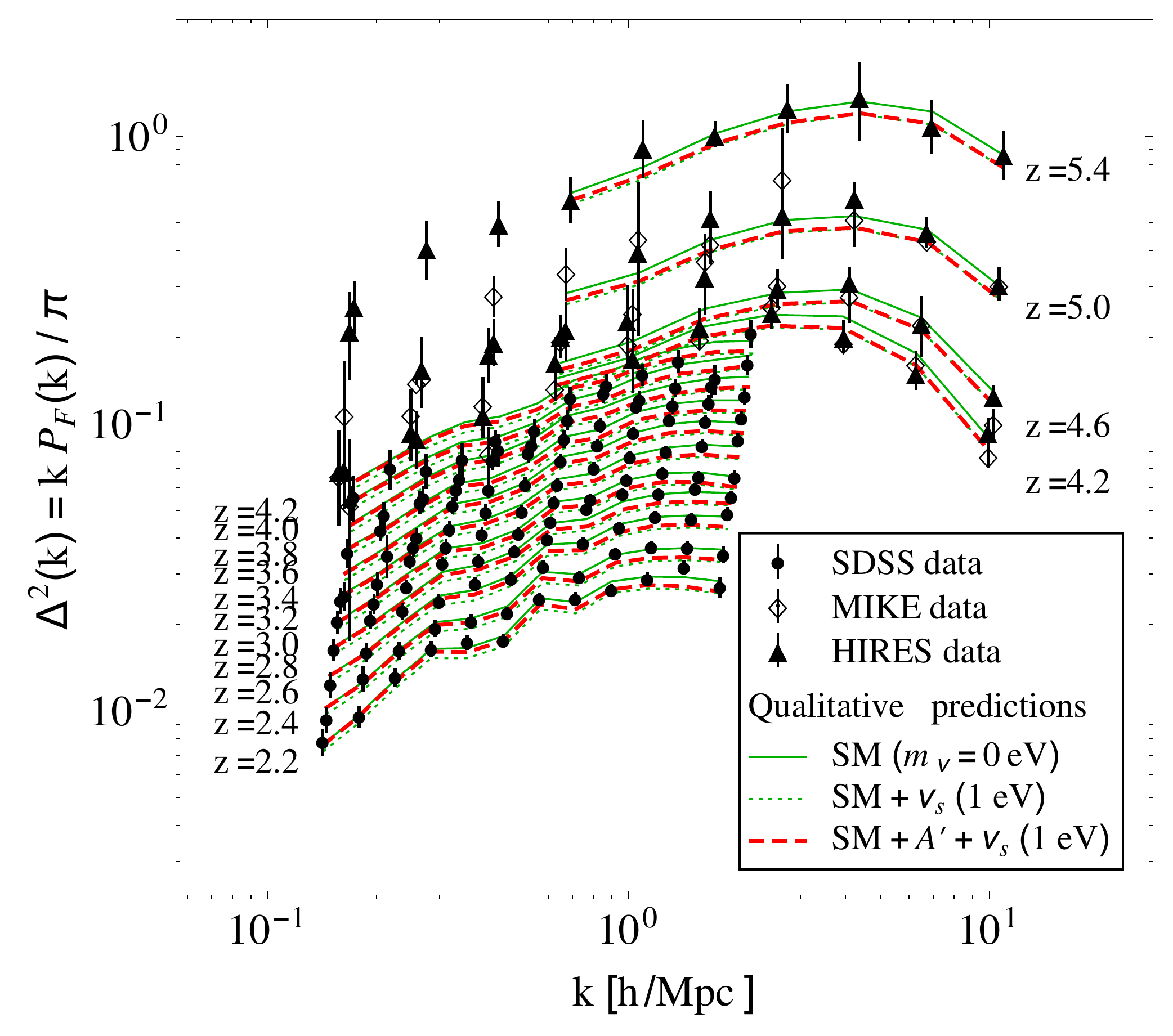} \\
    (a) & (b)
  \end{tabular}
  \caption{(a) Three-dimensional matter power spectrum $P_M(|\vec{k}|)$ derived
    from the SDSS Luminous Red Galaxy (LRG) Sample~\cite{Tegmark:2006az} and (b)
    one-dimensional flux power spectrum $\Delta^2(k) = k\,P_F(k) / \pi$ of Lyman-$\alpha$
    photons at various redshifts, compared to the qualitative predictions of
    sterile neutrino models with (red dashed curves) and without (green dotted
    curves) self-interactions.
    Note that the relation between $P_M(|\vec{k}|)$
    and $P_F(k)$ is non-linear, see e.g.~\cite{Rossi:2014wsa}.
    The assumed self-interaction parameters are $e_s = 0.1$,
    $M = 0.1$~MeV, and the assumed sterile neutrino mass is 1~eV.  The data points and
    the SM prediction (solid green curves) are taken from \cite{Tegmark:2006az}
    and and from~\cite{Viel:2013fqw}, respectively. The predictions including
    sterile neutrinos are obtained by multiplying the SM predictions by the
    $k$-dependent suppression profile from Fig.~7 of \cite{Lesgourgues:2012uu},
    shifted such that the onset of the suppression is at our calculated damping
    scale $k_s$ (Eq.(\ref{eq:ks}); see text for details), and scaled such that the maximum
    suppression is given by Eq.~\eqref{eq:DeltaP-lin} for panel (a) and by
    \eqref{eq:DeltaP-nonlin} for panel (b).  Note that the error bars
    shown here are statistical only, and large systematic uncertainties,
    especially at small scales (large $k$) should be kept in mind.}
  \label{fig:matter-power}
\end{figure*}

As a numerical example, for $M = 0.1\,\text{MeV}$, $e_s = 0.1$, we
find $\lambda_s^\text{coll} \simeq 29~\text{Mpc}/h$,
$\lambda^\text{fs}_s \simeq 68~\text{Mpc}/h$ and thus
\begin{align}
  \lambda_s \simeq 74~\text{Mpc}/h \,,
\end{align}
corresponding to a wave number of
\begin{align}
  k_s \equiv 2 \pi / \lambda_s  \simeq  0.085~h/\text{Mpc} \,.
\label{eq:ks}
\end{align}
This should be compared to the free streaming scale of a decoupled sterile
neutrino with a mass $\lesssim 1$~eV,
\begin{align}
  k_s^\text{no self-int.} \simeq 0.018 \, \sqrt{m\over \text{eV}} \,h/\text{Mpc} \,.
\end{align}
This factor of $\sim 5$ decrease in the free streaming scale compared to a
conventional sterile neutrinos without self-interactions implies that data on
large scale structure (LSS) and baryon acoustic oscillations (BAO) will be in
much better agreement with our model than with sterile neutrino models that do
not feature self-interactions. The strongest constraints will come from
data probing very small scales, in particular Lyman-$\alpha$ forests. 

Even at scales $k > k_s$, the suppression of the matter power spectrum
$P_M(|\vec{k}|)$ does not set in abruptly, but increases gradually.  For non-interacting
sterile neutrinos, numerical simulations show that the suppression saturates at
$k \simeq 50\, k_s$.  At even smaller scales (even larger $k$), the
deviation from the prediction of standard cosmology is
\cite{Lesgourgues:2006nd, Lesgourgues:2012uu}
\begin{align}
  \frac{\delta P_M(|\vec{k}|)}{P_M(|\vec{k}|)} \simeq -8 f_\nu
  \label{eq:DeltaP-lin}
\end{align}
in the linear structure formation regime.
Here, $f_\nu = 3 m_s \zeta(3) / (2\pi^2) \, T_s^3(t_0) \times 8\pi G / (3
H^2(t_0)) / \Omega_m \simeq 0.07$ is the ratio of the sterile neutrino mass
density $\Omega_s$ to the total mass density $\Omega_m \simeq 0.3$ today.
In the regime of non-linear structure formation, $\delta P_M(|\vec{k}|) / P_M(|\vec{k}|)$ is
somewhat larger \cite{Lesgourgues:2006nd, Lesgourgues:2012uu}, but
N-body simulations show that it
decreases again at scales $k \gtrsim \text{few}~h /
\text{Mpc}$~\cite{Brandbyge:2008rv, Rossi:2014wsa}.

It is, however,
difficult to directly measure $P_M(|\vec{k}|)$ at these nonlinear scales. The most sensitive
data sets are Lyman-$\alpha$ forests, from which the 1-dimensional \emph{flux} power spectrum
$P_F(k)$ of Lyman-$\alpha$ photons can be extracted. Translating
$P_F(k)$ into a measurement of $P_M(|\vec{k}|)$ requires a determination
of the bias $b(k)$, which is obtained from
numerical simulations of structure formation that include the dynamics of
the gas clouds in which Lyman-$\alpha$ photons from distant quasars are
absorbed. For SM neutrinos, such simulations have been performed for instance
in \cite{Rossi:2014wsa}, and we can estimate from Fig.~13 of that paper
that the maximal suppression of $P_F(k)$ is of order
\begin{align}
  \frac{\delta P_F(k)}{P_F(k)} \sim
    -0.1 \times \bigg( \frac{\sum m_\nu}{1\ \text{eV}}\bigg) \,,
  \label{eq:DeltaP-nonlin}
\end{align}
where $\sum m_\nu$ is the sum of all neutrino masses. This estimate is
crude but conservative, 
and ignores the fact that the maximal suppression
is actually smaller at lower redshifts. The suppression of
$P_F(k)$ described by Eq.~\eqref{eq:DeltaP-nonlin}
is \emph{smaller} than the suppression of $P_M(|\vec{k}|)$ from Eq.~\eqref{eq:DeltaP-lin}
because of the nonlinear $k$-dependent relation between the matter power spectrum and
the flux power spectrum (see for instance~\cite{Croft:2000hs}, especially
Fig.~16 in that paper).  Since no dedicated
simulations are available for our self-interacting sterile neutrino model,
we will assume in the following that $\delta P_F(k) / P_F(k)$
saturates at the value given by Eq.~\eqref{eq:DeltaP-nonlin} even
when $\sum m_\nu$ is dominated by the sterile neutrino mass $m_s$.
This amounts to assuming that the impact of secretly interacting sterile 
neutrinos on these small scales is \emph{qualitatively} similar to that of active neutrinos. 
A more detailed treatment requires a dedicated simulation including these secretly interacting 
sterile neutrinos.
Note that neutrino free-streaming after CMB decoupling may lead to less
suppression than described in
Eqs.~\eqref{eq:DeltaP-lin} and \eqref{eq:DeltaP-nonlin} because perturbation modes
well within the horizon have already grown significantly by that time.  We will
not include this effect
in the following discussion to remain conservative.

We show the qualitative impact of self-interacting sterile neutrinos on large
scale structure in Fig.~\ref{fig:matter-power}. Panel (a) compares theoretical
predictions in models with and without sterile neutrinos to data on the
three-dimensional matter power spectrum $P_M(|\vec{k}|)$ from the Sloan Digital
Sky Survey (SDSS) Luminous Red Galaxy (LRG) catalog~\cite{Tegmark:2006az}.
Panel (b) compares to one-dimensional flux power spectra $\Delta^2(k) \equiv k
\, P_F(k) / \pi$ from Lyman-$\alpha$ forest data~\cite{Viel:2013fqw}.  SDSS-LRG data
corresponds to a mean redshift of $z \simeq 0.35$, while Lyman-$\alpha$ data is
split up according to redshift and reaches up to $z \simeq 5.4$.  Note that the
data in \cite{Viel:2013fqw} is presented as a function of the wave number $k_v$
in velocity space, measured in units of sec/km.  The conversion to the wave
number $k$ in coordinate space, measured in units of $h/\text{Mpc}$, is done
according to the formula $k = k_v \, H(z) / (1+z)$, where $H(z)$ is the Hubble
rate at redshift $z$. The theoretical predictions for the SM with vanishing
neutrino mass (solid green curves in Fig.~\ref{fig:matter-power}) are taken
from \cite{Tegmark:2006az} and \cite{Viel:2013fqw}, respectively. Our
(qualitative) predictions for sterile neutrino models with and without
self-interactions are obtained in the following way: we start from the
numerical prediction for the neutrino-induced suppression of the matter power
spectrum from Ref.~\cite{Lesgourgues:2012uu}.  In particular, we use the curve
corresponding to $f_\nu = 0.07$ from Fig.~7 in that paper. We then shift this
curve such that the onset of the suppression coincides with our calculated
damping scale $k_s$, and we rescale it such that the maximal suppression
is $-8 f_\nu$ in Fig.~\ref{fig:matter-power} (a) (linear regime) and 10\%
in Fig.~\ref{fig:matter-power} (b) (nonlinear regime), see Eqs.~\eqref{eq:DeltaP-lin}
and \eqref{eq:DeltaP-nonlin}.  We then multiply
with the SM prediction to obtain the dotted green curves for sterile neutrinos
without self-interactions and the red dashed curves for sterile neutrinos with
self-interactions in Fig.~\ref{fig:matter-power}. We use $e_s = 0.1$, $M =
0.1$~MeV and $m_s = 1$~eV.  Since we neglect a possible upturn of the
power spectrum at $k \gtrsim 1~h / \text{Mpc}$~\cite{Brandbyge:2008rv},
our estimates are very conservative.

From Fig.~\ref{fig:matter-power} (a), we observe that the suppression of the
matter power spectrum at scales $\lesssim 0.2~h$/Mpc due to self-interacting
sterile neutrinos is completely negligible, while a fully thermalized
non-interacting sterile neutrino with the same mass leads to a clear
suppression already at these scales.  This implies that self-interacting
sterile neutrinos with the parameters chosen here are not constrained by data
on linear structure formation.  Going to smaller scales or larger $k$
(Fig.~\ref{fig:matter-power} (b)), where non-linear effects become relevant, we
see that both the sterile neutrino model with self-interactions and the one
without lead to suppression, but the amount of suppression is reduced in
the self-interacting case.  It was shown in Ref.~\cite{Viel:2013fqw} that the
data disfavors suppression larger than 10\% at $k=10 \,h/\text{Mpc}$.
Self-interacting sterile neutrinos at the benchmark point shown in
Fig.~\ref{fig:matter-power} appear to be marginally consistent with
this constraint. It should be
kept in mind, however, that our predictions are only qualitative.
Therefore, only a detailed fit using
simulations of non-linear structure formation that include sterile neutrino
self-interactions could provide a conclusive assessment of the viability of
such a scenario.

Let us finally discuss how the cosmological effects of the three active neutrinos
are modified in the self-interacting sterile neutrino
scenario. The dynamics of the mass eigenstates $\nu_2$ and $\nu_3$, which we
assume not to mix with $\nu_4$, is the same as in standard cosmology: they
start to free stream at redshift $z \gg 10^5$.  $\nu_1$, however, starts to
free stream later than a non-interacting neutrino, but earlier than $\nu_4$.
The free-streaming condition for $\nu_1$ is, in analogy to
Eq.~\eqref{eq:fs-cond},
\begin{align}
  (T_\nu^\text{SM})^3 \cdot (T_\nu^\text{SM})^2 \bigg({e_s^2 \over M^2}\bigg)^2
    \sin^2\theta_0 \lesssim H \,.
  \label{eq:fs-cond-nu-1}
\end{align}
As shown in Ref.~\cite{Cyr-Racine:2013jua}, free-streaming of active neutrinos
before redshift $z \sim 10^5$ is required to sufficiently suppress the acoustic
peaks in the CMB power spectrum.  The change from three to two truly free streaming
neutrino species in our model will lead to minor modifications of the CMB power spectrum,
but the analysis from \cite{Cyr-Racine:2013jua} suggests that these are unlikely
to spoil the fit to CMB data, in particular since they may be compensated by
changes in the best fit values of other cosmological parameters.

Note that also Planck CMB data alone, without including data on large scale
structure observations, imposes an upper limit on the mass of sterile
neutrinos, which, for a fully thermalized species is $m_s \lesssim 0.5$~eV at
95\% C.L.~\cite{Planck:2015xua}.  A much weaker bound is expected 
if self- interactions among sterile neutrinos are so strong that they remain 
collisional until after the CMB epoch. In this case, the early Integrated Sachs-Wolfe (ISW) effect induced by $\nu_s$
perturbations at low multipole order ($50\le l \le 200$) will be suppressed
\cite{Lesgourgues:2012uu}. Thus, the main effect on the CMB will come from the
shift of matter--radiation equality, to which the sensitivity is, however, much
weaker.

\section{Discussion and conclusions}
\label{sec:conclusions}

As we have seen in the previous section, there are two main scenarios in which
self-interacting sterile neutrinos do not run into conflict with cosmological data:

{\it (i)} The $\nu_s$ production rate $\Gamma_s$ drops below the Hubble expansion
rate $H$ before the effective potential $|V_\text{eff}|$ drops below the
oscillation frequency $|\Delta m^2/(2E)|$ and the dynamic suppression of
active--sterile mixing due to $V_\text{eff}$ ends. In this case, sterile
neutrinos are not produced in significant numbers in the early Universe and
hence cosmology is not sensitive to their existence.  An explanation of small
scale structure anomalies as advocated in~\cite{Dasgupta:2013zpn} is, however,
not possible in this scenario.\footnote{Note that recent simulations
of cosmological structure formation suggest that these anomalies may be
resolved once baryons are included in the
simulations~\cite{Vogelsberger:2014kha,Sawala:2014xka}.}  In particular, even
if the new interaction also couples to dark matter as proposed
in~\cite{Dasgupta:2013zpn}, it is too weak to have phenomenological
consequences.

This disadvantage can be avoided if more than one self-interacting sterile
neutrino exists. Consider for example, a model with three mostly sterile
neutrino mass eigenstates $\nu_4$, $\nu_5$, $\nu_6$.  Let $\nu_4$ and $\nu_5$
have a relatively large mixing $\theta_0 \sim 0.1$ with active neutrinos, as
motivated for instance by the short baseline oscillation anomalies. Let their
coupling to the $A'$ gauge boson be $e_s^{(4,5)} \simeq 10^{-5}$, large
enough to dynamically suppress their mixing with the mostly active mass
eigenstates until after BBN, but small enough to prevent their equilibration
afterwards.  On the other hand, let $\nu_6$ have a vanishing mixing with
$\nu_{1,2,3}$, but a larger secret gauge coupling $e_s^{(6)} \simeq
0.1$.  Due to its small mixing, it is never produced through oscillations.
However, its primordial population---the relic density produced before the
visible and sterile sectors decoupled in the very early Universe---still
acts as a thermal bath to which the dark matter may be strongly coupled, thus
potentially solving the missing satellites problem~\cite{Klypin:1999uc,
Boehm:2000gq, Bringmann:2006mu, Aarssen:2012fx, Shoemaker:2013tda}.

{\it (ii)} The self-interactions are so strong that sterile neutrinos remain
collisional at least until matter--radiation equality.  In this scenario,
sterile neutrinos are produced when $|V_\text{eff}| \leq |\Delta m^2/(2E)|$.
However, as shown in Sec.~\ref{sec:Neff}, the effective number of relativistic
degrees of freedom in the Universe, $N_\text{eff}$, remains close to 3 because
equilibration between active and sterile neutrinos happens after neutrinos have
decoupled from the photon bath.  Moreover, as argued in Sec.~\ref{sec:LSS}, the
impact of self-interacting sterile neutrinos on structure formation is much
smaller in this scenario than the impact of conventional non-interacting
sterile neutrino because they cannot transport energy efficiently over large
distances due to their reduced free-streaming. Structure formation constraints
could be further relaxed in models containing, besides an eV-scale mass eigenstate
$\nu_4$, one or more additional mostly sterile states with much lower
masses~\cite{Tang:2014yla}.
It is intriguing that the parameter region corresponding to scenario
{\it (ii)} contains the region where small scale structure anomalies can be
explained, as shown in~\cite{Dasgupta:2013zpn}.

\begin{figure}
  \centering
  \includegraphics[width=0.9\columnwidth]{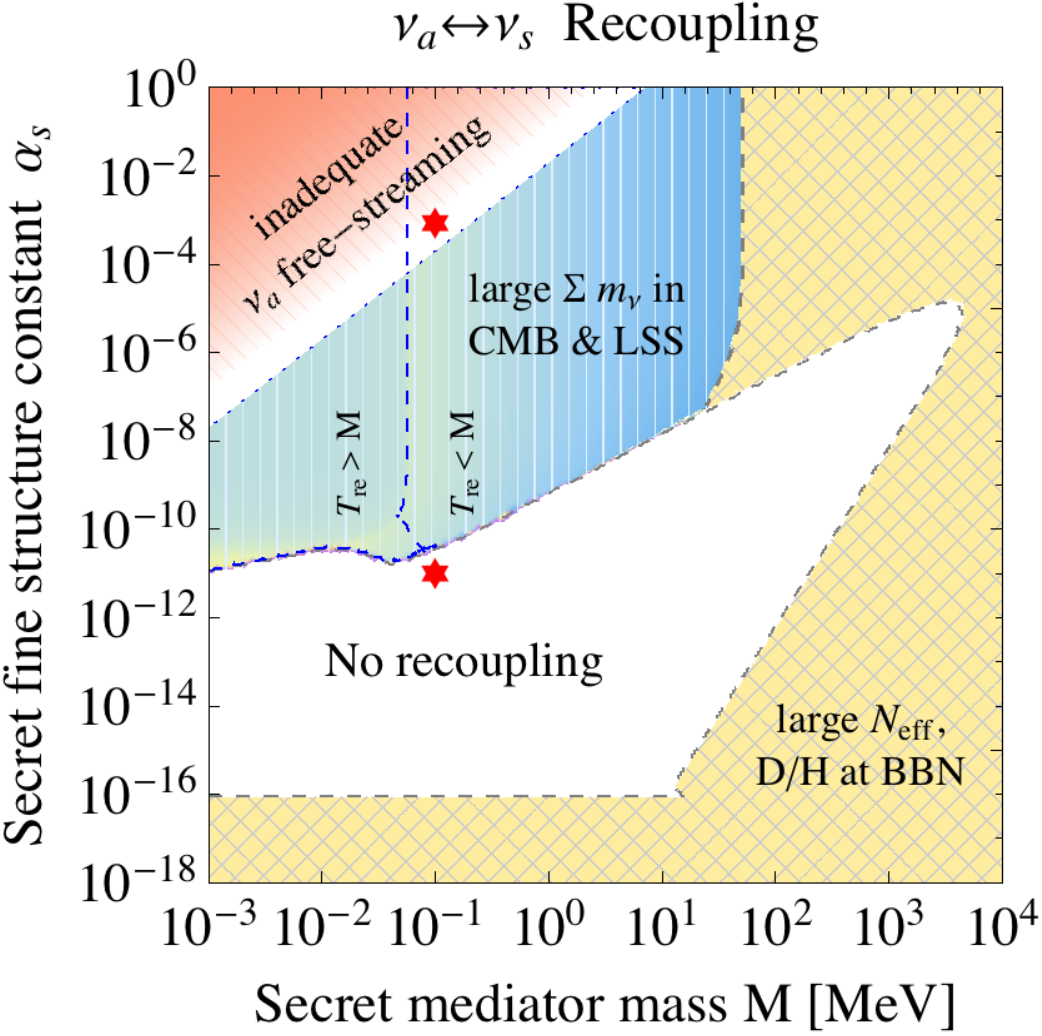}
  \caption{Schematic illustration of the parameter space for eV-scale sterile
    neutrinos coupled to a new ``secret'' gauge boson with mass $M$ and a
    secret fine structure constant $\alpha_s$.  The vacuum mixing angle between
    active and sterile neutrinos was taken to be $\theta_0 = 0.1$.
    The white region in the lower half of the plot is allowed by all constraints,
    while the narrow white band in the upper left part satisfies all constraints
    except possibly large scale structure (LSS)
    limits from Lyman-$\alpha$ data at the smallest scales. The red stars show
    representative models in scenarios {\it (i)} and {\it (ii)}.
    The colored regions are excluded, either by LSS observations (blue vertically
    hatched), by the requirement that active neutrinos should free stream
    early enough (red shaded), or by a combination of CMB and BBN data
    (yellow cross-hatched).}
  \label{fig:paramspace}
\end{figure}

We summarize these results in Fig.~\ref{fig:paramspace}. The yellow
cross-hatched region on the right is ruled out because active and sterile
neutrinos come into thermal equilibrium before the active neutrino decoupling
from the SM plasma. In the lower part of this region, this happens simply
because $V_\text{eff}$ is negligibly small.  In the upper part, $V_\text{eff}$
is large, but also $\Gamma_s$ is large so that collisional production of
sterile neutrinos is efficient in spite of the suppressed in-medium mixing
$\theta_m$.  This leads to constraints from $N_\text{eff}$ and from the light
element abundances in BBN~\cite{Saviano:2014esa}.  In the blue vertically
hatched region, sterile neutrinos are produced after $\nu_a$ decoupling, so
that CMB constraints on $N_\text{eff}$ remain satisfied. However, sterile
neutrinos free-stream early on in this region and violate the CMB and structure
formation constraints on their mass.  This mass constraint can be considerably
relaxed if the sterile neutrinos remain collisional until \emph{after} the CMB
epoch at $T_\gamma \simeq 0.3$~eV. This defines the upper edge of the blue
hatched region. In the red shaded region in the upper left corner, the secret
interaction is too strong and $\nu_1$ free streams too late. CMB data requires
that active neutrinos free stream early enough, and thus strongly disfavors this
region.  Two white regions remain allowed: Scenario $(i)$ with weak
self-interactions, corresponds to the wedge-shaped white region in the lower
part of the plot. Scenario $(ii)$, with strong self-interactions, is
realized in the thin white band between the blue vertically hatched region and
the red shaded region. As explained above, whether or not this white band
is allowed depends strongly on systematic uncertainties at Lyman-$\alpha$
scales and on the possible existence of additional states with masses $\ll 1$~eV.

There are several important caveats and limitations to the above analysis.
First, we have only worked with thermal averages for the parameters
characterizing each particle species, such as energy, velocity, pressure, etc.
To obtain more accurate predictions, it would be necessary to solve
momentum-dependent quantum kinetic equations.  This would be in particular
interesting in the temperature regions where $V_\text{eff}$ changes sign
and where $V_\text{eff} \sim \Delta m^2 / (2 T_{\nu_a}) \times \cos 2\theta_0$.
We expect that our modeling of flavor conversions in this region as fully
non-adiabatic transitions is accurate, but this assumption remains to be
checked explicitly.  Moreover, the impact of self-interacting sterile neutrinos
on non-linear structure formation at the smallest scales probed by
Lyman-$\alpha$ data should be calculated more carefully.
Improving these issues is left for future work.

In conclusion, we have argued in this paper that self-interacting sterile
neutrinos remain a cosmologically viable extension of the Standard Model. As
long as the self-interaction dynamically suppresses sterile neutrino production
until neutrinos decouple from the photon bath, the abundance produced
afterwards is not in conflict with constraints on $N_\text{eff}$.  Moreover, if
the self-interaction is either weak enough for scattering to be negligible
after the dynamic mixing suppression is lifted, or strong enough to delay free
streaming of sterile neutrinos until sufficiently late times, also structure
formation constraints can be avoided or significantly relaxed.

\section*{Acknowledgments}

We are grateful to Vid Irsic, Gianpiero Mangano, Alessandro Mirizzi and Ninetta Saviano for
very useful discussions. Moreover, it is a pleasure to thank Matteo Viel for
providing the Lyman-$\alpha$ data underlying Fig.~\ref{fig:matter-power} (b) in
machine-readable form and for discussing it with us.

\appendix
\section{Kinetic temperature and pressure}
\label{sec:Tkin}

In the following, we give more details on the momentum
distribution function $f(p,t)$ of sterile neutrinos $\nu_s$ after they have
decoupled from all other particle species. $f(p,t)$ is essential in the
calculation of the pressure $P$ in Sec.~\ref{sec:Neff} and the
average velocity $\ev{v_s}$ in Sec.~\ref{sec:LSS}.

Even when $\nu_s$ are decoupled from other
particles, they may still couple to themselves via strong self-interactions.
If the self-interaction conserves the number of particles, such as
$\nu_s \nu_s \leftrightarrow \nu_s \nu_s$, it only maintains kinetic
equilibrium, but not chemical equilibrium. Number conservation
and entropy maximization force the $\nu_s$ momentum distribution
function in kinetic equilibrium to take the form
\begin{align}
  f(p, t) = \frac{1}{e^{[E(p) - \mu_s(t)]/T_{s}(t)} + 1} \,,
  \label{kinetical:MDF}
\end{align}
where $T_{s}(t)$ is defined as the \emph{kinetic temperature},
$\mu_s(t)$ is the chemical potential, and $E(p) = (p^2 + m_s^2)^{1/2}$.
Here and in the following, we use the definition $p \equiv |\vec{p}|$.
Since we are interested in the evolution at relatively late times,
when the sterile neutrino density is low compared to the density of a
degenerate fermion gas and thus $\ev{f(p,t)} \ll 1$, the classical
approximation
\begin{align}
  f(p,t) \simeq e^{-[E(p) - \mu_s(t)]/T_{s}(t)}
\end{align}
is adequate.

Our goal is to solve for the functions $T_{s}(t)$ and $\mu_s(t)$
with the initial condition of a relativistic thermal ensemble of
sterile neutrinos. This means that, initially, $T_s =T_i \gg m_s$
and $\mu_s = 0$ at $a=a_i$.
Note that the sterile neutrino mass will lead to
a non-zero $\mu_s$ soon after neutrinos go out of chemical 
equilibrium. Although it is difficult to analytically
solve the corresponding Boltzmann equation, there are two conditions that
can be used to numerically obtain $T_{s}$ and $\mu_s$ as functions of the scale
factor $a(t)$. One is number conservation. The other is entropy
conservation, which holds approximately for kinetic equilibrium
in the classical limit \cite{Bernstein:1988bw}.

\begin{figure}
  \centering
  \includegraphics[width=0.85\columnwidth]{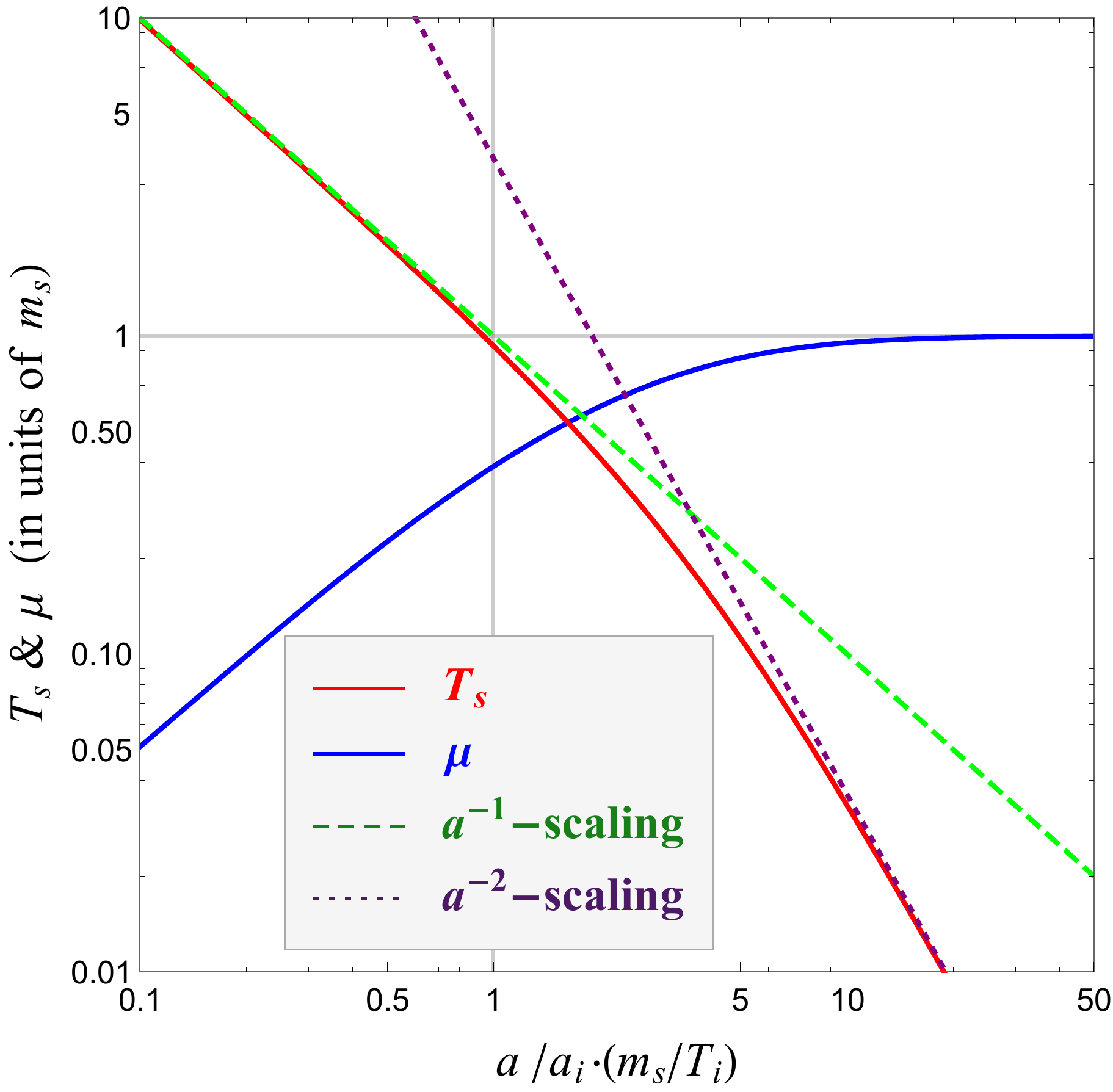}
  \caption{Kinetic temperature $T_s$ and chemical potential $\mu_s$ of sterile
    neutrinos, as functions of the scale factor $a(t)$ during the transition from
    the relativistic regime to the non-relativistic regime, with initial conditions
    $T_s = T_i \gg m_s$ and $\mu_s = 0$ at $a=a_i$.}
  \label{fig:pressure}
  \vspace{-.4cm}
\end{figure}

The number density is
\begin{align}
  n_s(t) &= \int \! \frac{d^3 p}{(2\pi)^3} \, f(p, t)
\end{align}
and the classical entropy density is defined as 
\begin{align}
  s_s(t) \equiv \int \! \frac{d^3 p}{(2\pi)^3} \, f(p,t) \, [1 - \ln f(p,t)] \,.
\end{align}
It is straightforward to obtain the asymptotic solutions~\cite{Bernstein:1988bw}
\begin{align}
  T_{s}(t) &\propto
     \begin{cases}
       a^{-1}(t) & \text{for $T_s \gg m_s$} \\[0.5em]
       a^{-2}(t) & \text{for $T_s \ll m_s$}
     \end{cases} \, \\
\intertext{and}
  \mu_s(t) &\propto
     \begin{cases}
       a(t)         & \text{for $T_s \gg m_s$} \\[0.5em]
       \text{const} & \text{for $T_s \ll m_s$}
     \end{cases} \,.
\end{align}
In the transition region $T_s \sim m_s$, the solution needs to be obtained
numerically. The result is plotted in Fig.~\ref{fig:pressure}.

Finally, we comment on the calculation of the pressure $P$ and the average
velocity $\ev{v_s}$ of sterile neutrinos.  The pressure is given by \cite{Bernstein:1988bw}
\begin{align}
  P &\equiv \int \! \frac{d^3 p}{(2\pi)^3} \, \frac{p^2}{3E} \, f(p,t)
                                                            \nonumber\\[1ex]
       &= -T_s \, e^{\mu_s/T_s} \int \! \frac{d^3 p}{(2\pi)^3} \, \frac{p}{3} \, \frac{d}{dp}
               e^{-E(p)/T_{s}}
                                                            \nonumber\\[1ex]
       &= T_{s} \cdot n_s \,,
\end{align}
and the average velocity is
\begin{align}
  \ev{v_s} \simeq
    \frac{1}{N} \int\!\frac{d^3 p}{(2\pi)^3} \frac{p}{E(p)} \, f(p, t) \,.
\end{align}
Here, $N \equiv \int\!dp\,4\pi p^2 / (2\pi)^3 \times f(p, t)$ is a normalization factor.
Besides the conditions of kinetic
equilibrium discussed above, we also need to take into account that sterile
neutrino self-interactions freeze out at a time $t^\text{dec}$ and sterile
sector temperature $T_s = T_{s,\text{dec}}$,
after which kinetic equilibrium is lost and sterile neutrino
momenta are simply redshifted as
$a^{-1}(t)$. This implies for the momentum distribution function:
\begin{widetext}
\begin{align}
  f(p, t) = \!
    \begin{cases}
      \frac{1}{\exp\!\big[
                 \frac{1}{T_s(t)} \big( \sqrt{p^2 + m_s^2} - \mu_s(t) \big) \big] + 1}
            &\text{for $t < t^\text{dec}$} \\[0.4cm]
      \frac{1}{\exp\!\big[
                 \frac{1}{T_{s,\text{dec}}} \big( \sqrt{\frac{p^2 a^2(t)}{a^2(t^\text{dec})}
                   + m_s^2} - \mu_s(t^\text{dec}) \big) \big] + 1}
            &\text{for $t > t^\text{dec}$}
    \end{cases} \,.
\end{align}
\end{widetext}
Here, $\mu_s(t^\text{dec})$ is the chemical potential at the time of decoupling.
We have checked that the exact decoupling time only slightly changes the evolution of
$P$, so we regard our solution in Fig.~\ref{fig:pressure} as universal for all
parameter values of interest. In Sec.~\ref{sec:LSS}, we have for simplicity
assumed a sudden decoupling of self-interactions though. For the value $e_s^2/M^2 \simeq
1$~MeV$^{-2}$ chosen there, this leads to $T_{s,\text{dec}} \sim 0.0024$~eV,
corresponding to a photon temperature of 0.038~eV.

\bibliographystyle{JHEP}
\bibliography{reference}

\end{document}